\documentclass{pasj01}
\pdfoutput=1
\Received{12-Oct-2023}
\Accepted{18-Dec-2023}

\newcommand{\Ha}{\mathrm{H}\mathrm{\alpha}}

\newcommand{\Ntwo}{\mathrm{N}\mathrm{I}\hspace{-1.2pt}\mathrm{I}}
\newcommand{\Stwo}{\mathrm{S}\mathrm{I}\hspace{-1.2pt}\mathrm{I}}
\newcommand{\Oone}{\mathrm{O}\mathrm{I}}

 \newcommand{\sdss}{SDSS J1430+2303\ }
 \newcommand{\mrm}{\ \mathrm}
 \newcommand{\AAA}{\ \mathrm{\mathring{A}}}
 
 \newcommand{\BLone}{\mathrm{BL}_1}
 \newcommand{\BLtwo}{\mathrm{BL}_2}
\usepackage{graphicx}
\usepackage{epstopdf}
 \usepackage[switch,mathlines]{lineno}
 
\begin{document}

\title{ 
The Variability of the Broad Line Profiles of SDSS J1430+2303 }
\author{Atsushi \textsc{Hoshi}\altaffilmark{1,2}%
\thanks{Example: Present Address is xxxxxxxxxx}}
\altaffiltext{1}{Astronomical Institute, Tohoku University, 6-3 Aramaki, Aoba-ku, Sendai, Miyagi 980-8578, Japan}
\email{hoshi@astr.tohoku.ac.jp}

\author{Toru \textsc{Yamada},\altaffilmark{2,1}}
\altaffiltext{2}{Institute of Space and Astronautical Science, Japan Aerospace Exploration Agency, 3-1-1, Yoshinodai, Chuou-ku, Sagamihara, Kanagawa, 252-5210, Japan}

\author{Kouji \textsc{Ohta}\altaffilmark{3}}
\altaffiltext{3}{Department of Astronomy, Kyoto University, Kitashirakawa-Oiwake-cho, Sakyou-ku, Kyoto, 606-8502, Japan}

\maketitle

\begin{abstract}
SDSS J1430+2303 has been argued to possess a supermassive black hole binary which is predicted to merge within a few months or three years from January 2022.
We conducted follow-up optical spectroscopic observations of SDSS J1430+2303 with KOOLS-IFU on Seimei Telescope in May, June, and July 2022, and April 2023.
The observed spectrum around $\mathrm{H}\mathrm{\alpha}$ shows a central broad component $\sim 10^3\ \mathrm{km\ s^{-1}}$ blueshifted from the narrow H$\mathrm{\alpha}$ line as well as the broader double-peaked component with a separation of $\sim\pm 5\times10^3\ \mathrm{km\ s^{-1}}$, similar to the spectrum reported in January 2022.
We investigate the variability of the complex broad $\mathrm{H}\mathrm{\alpha}$ emission line relative to the continuum over the observation period. 
The continuum-normalized relative flux of the central broad component shows the increasing trend from May to July 2022 which is interpreted to be caused by the decrease of the continuum as also supported by damping of the X-ray, UV, and optical light curves observed for the same period. 
From July 2022 to April 2023, however, the central broad component decreased significantly.
For  the relative flux of the broader double-peaked component, on the other hand, no significant change appears at any epoch.
These results suggest that the complicated broad line profile of SDSS J1430+2303 is generated from at least two distinct regions.
While the central broad component originates from a broad line region, the broader double-peaked component arises in the vicinity of the continuum source.
\end{abstract}

\section{Introduction}
\begin{table*}[htbp]
 
 \caption{Summary of the Observations}
 \centering
  \label{observation}
   \begin{tabular}{cccccccc}
    \hline \hline
  Epoch&Proposal ID&  Date&MJD&From Jan 1, 2022&Grism&Total Exposure&Standard star
    \\
   \hline
(a)&22A-K-0029&May 27, 2022 &59726&146 days& VPH-blue   & 1800 sec&HZ 44 \\
 (b)&22A-K-0029&June 18, 2022 &59748&168 days& VPH683O56   & 3600 sec& BD+33$^\circ$2642 \\
   (c)&22B-N-CN05&July 28, 2022 &59788&208 days& VPH-blue   & 1800 sec& BD+33$^\circ$2642 \\
   (d)&23A-N-CT01&April 17, 2023 &60051&471 days& VPH-blue   & 1800 sec& BD+33$^\circ$2642 \\
\hline
  \end{tabular}
 \end{table*}
 
Active galactic nuclei (AGNs) play an important role in understanding the growth of supermassive black holes (SMBHs) and their relationship with the host galaxies. A crucial question is how SMBHs formed and evolved in the hierarchical growth of galaxies in the universe \citep{silkrees1998,hopkins2008cosmological,kormendy2013coevolution}.
While SMBHs are formed in galaxies and they increase their mass by the subsequent gas accretion events, they can also grow by merging of the galaxies which already host the SMBH \citep{rees1984black}.\\
More than a few supermassive black hole binary (SMBHB) systems and candidates have been identified and extensively studied so far \citep{begelman1980massive,roos1981galaxy,de2019quest,sillanpaa1988oj,bhatta2016detectionoj287,oj287dey2019unique,komossa2021supermassive}.
They are observationally characterized by the resolved close pair of AGN or more typically the existence of double-peaked emission lines, and sometimes by the distinctive quasi-periodic photometric light curves \citep{shen2010identifying}.
The orbital motion of SMBHB may also affect their broad emission line profiles and the cases that the broad lines significantly shifted from the narrow lines were observed \citep{shen2010identifying,popovic2012super}.\\
SDSS J143016.05+230344.4 (hereafter \sdss, also known as AT2019cuk and Tick Tock) which is a Seyfert1 galaxy at redshift 0.08105 with AB magnitude $\sim16.5$ in $r\mathchar`-$band, is argued to possess the SMBHB system just before the coalescence \citep{jiang2022tick}. The rapid decay of the amplitudes and periods of the light curves in X-ray, UV and optical wavelength observed by January 2022 predicts that SMBHB in \sdss can coalesce within $\sim100$ days or three years \citep{jiang2022tick}. 
On the other hands, a recent follow-up observations argued that the optical light curve is hardly explained by the binary scenario \citep{dotti2023optical}.
It is interesting to note that the optical broad emission line profile of this object also dramatically changed between 2005 and 2022.
While the spectrum of \sdss taken in 2005 showed only one broad line which is blueshifted ($\sim2400\ \mathrm{km\ s^{-1}}$) from the narrow $\Ha$ line, the spectrum taken in January 2022 showed the drastically changed puzzling profiles of a broad line which is blueshifted ($\sim900\ \mathrm{km\ s^{-1}}$) from the narrow $\Ha$ together with the component which is represented by a pair of two broad lines with symmetry \citep{jiang2022tick}.
Such abrupt changes in the broad line profile are considered quite unique even among the AGNs with the most prominent line profile variations and it's important to monitor the phenomena over a sufficiently long time scale, whether or not it is directly related to the predicted SMBHB coalescence itself.\\
A subclass of AGNs is also known to show changes in their broad emission lines in the UV/optical spectrum and called as the changing-look (or changing-state) AGNs \citep{macleod2016systematic,stern2018mid,macleod2019changing,ross2020first,ricci2023changing}. Although the physical origins of changing-look AGNs are still under debate, the phenomena can be related to the change in the accretion rate \citep{graham2020understanding}.
While \sdss can be broadly categorized as a changing-look AGN, the origin of the broad line variation may differ from that of changing-look AGNs where typically a single broad emission appears or disappears and such drastic change of the velocity profile in \sdss is rarely observed.

In this paper, we report our results of optical spectroscopy of \sdss at the $\Ha$ wavelength region over the four epochs from May 2022 to April 2023. We present that the overall line structure is similar to that observed in January 2022 but there is a significant change in a part of the broad components.\\
Throughout this paper, we assume a $\mathrm{\Lambda CDM}$ cosmological parameters of $\Omega_m=0.3$ and $\Omega_\Lambda =0.7$ and Hubble constant $H_0=70\mrm{km\ s^{-1} Mpc^{-1}}$. We use the AB magnitude system.

\section{Observations and Data Reduction}
\begin{figure*}[ht]
 \hspace{6mm}
 \includegraphics[width=156mm]{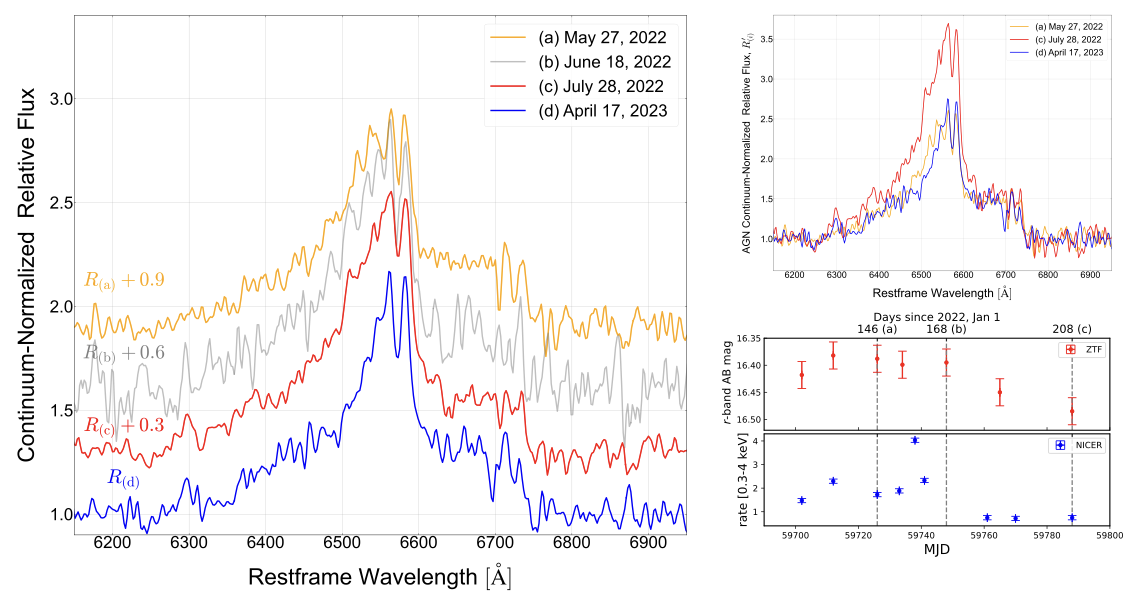}
\caption{(left):Continuum-normalized spectra $R_{(a)}+0.9$, $R_{(b)}+0.6$, $R_{(c)}+0.3$, and $R_{(d)}$. The spectra observed in epoch (a), (b), (c) and (d) correspond to orange, gray, red and blue lines. (right): The AGN continuum-normalized spectra $R^{\prime}_{(a)}$, $R^{\prime}_{(c)}$ and $R^{\prime}_{(d)}$ after subtracting the host component are shown in upper panel. In epoch (b), the host component is not measured due to the limited wavelength coverage. The light curves of \sdss in $r$-band and X-ray at around the observed epochs referring to \citet{masterson2023unusual} are shown in the lower panel.}
\label{data}
\end{figure*}

\sdss was observed with KOOLS-IFU which is a fiber-fed integral field spectrograph equipped with the 3.8 m Seimei telescope \citep{matsubayashi2019kools, kurita2020seimei}. 
The instrument consists of the 110 fibers for objects (after October 2020), which achieves a total FoV of $8.4\times8.0$ arcsec$^2$.
We conducted the spectroscopic observations on May 27, June 18 (Proposal ID : 22A-K-0029, PI : Ohta) and July 28, 2022(Proposal ID : 22B-N-CN05 PI : Hoshi), and April 17, 2023 (Proposal ID : 23A-N-CT01 PI : Hoshi). These observation epochs are denoted as (a), (b), (c) and (d).
The information of the observations is summarized in Table \ref{observation}.
We used the VPH-blue (4100-8900$\AAA$) and VPH683O56 (6200–7500$\AAA$) grisms. 
The spectral resolution ($R =\lambda/\Delta \lambda$) of VPH-blue and VPH683O56 are $R\sim500$ and $R\sim 2000$, respectively. 
\\
The data reduction such as spectrum extraction, flat fielding, and wavelength calibration was performed by using {\tt\string PyRAF}, a python package developed by Space Telescope Science Institute (STScI) using the scripts developed for KOOLS-IFU. 
Spectroscopic standard stars, HZ44 and BD+33$^\circ$2642 were observed for relative flux calibration over the observed wavelength ranges \citep{oke1990faint}.
We also corrected the atmospheric absorption around the redshifted $\Ha$ lines. For all the observation epochs, flat-fielding frames were created by using the twilight sky and the wavelength calibration was performed by using the arc lamps (Hg, Ne, Xe). The wavelength sampling is typically $\sim 2\AAA$ per pixel and the wavelength determination accuracy is $\sim0.25\AAA$. Median stacking was used to remove cosmic rays.

\section{Results and Discussion}
\subsection{Broad Line Decomposition}
In order to investigate the variability of the broad $\Ha$ emission line, we present the continuum-normalized spectra in Figure \ref{data} (left). 
Using the \texttt{curve\_fit} function from the \texttt{scipy} module \citep{virtanen2020scipy} in Python, we determined the relative flux ($R_{(i)}$), which is normalized by the continuum. 
The index $i$ corresponds to each epoch (a, b, c, d). 
The continuum which includes both the AGN and the host galaxy components was derived by fitting the spectrum within the wavelength ranges of 6150-6250$\AAA$ and 6850-6950$\AAA$ using a linear function.
We also show the AGN continuum-normalized spectra ($R^{\prime}_{(i)}$) after subtracting the host component in Figure \ref{data} (upper right).
Since \sdss is classified as a typical elliptical galaxy \citep{jiang2022tick}, the host contribution is estimated by the equivalent width (EW) of Na D absorption line using a template spectrum of the elliptical galaxy in \citet{kinney1996template}.
Although the Na D absorption may suffer from interstellar absorption, its wavelength is close to that of $\Ha$ and we here ignored the Na D interstellar absorption, which corresponds to the case of maximum contribution of the host component.
Unfortunately, we are not able to estimate the fraction of the host galaxy in epoch (b) due to the limited wavelength coverage of VPH683O56 grism.
The lower panel in Figure \ref{data} shows the light curves in $r$-band and X-ray during the observation period up to epoch (c) from \citet{masterson2023unusual}.

The spectra in the $\Ha$ region exhibit a consistently complicated profile over all epochs. 
To decompose the possible multiple components we fit the spectra using a linear continuum in the range 6100-7000$\AAA$ and simple multiple Gaussian components in the $\Ha$ region.
The velocity widths of the five narrow emission lines, namely, $\Ha$, $[\Ntwo]\lambda\lambda 6548, 6583$, and $\ [\Stwo]\lambda\lambda 6717, 6734$ are fixed at the same value for each epoch. 
The flux ratio $\frac{[\Ntwo]\lambda 6583}{[\Ntwo]\lambda 6548}=4.26$ is fixed with the typical composite quasar spectra from SDSS \citep{composit}. 
The $[\Oone]\lambda 6300$ line is also fitted simultaneously, but the velocity width is not fixed.

\begin{figure}[ht]

 \includegraphics[width=77mm]{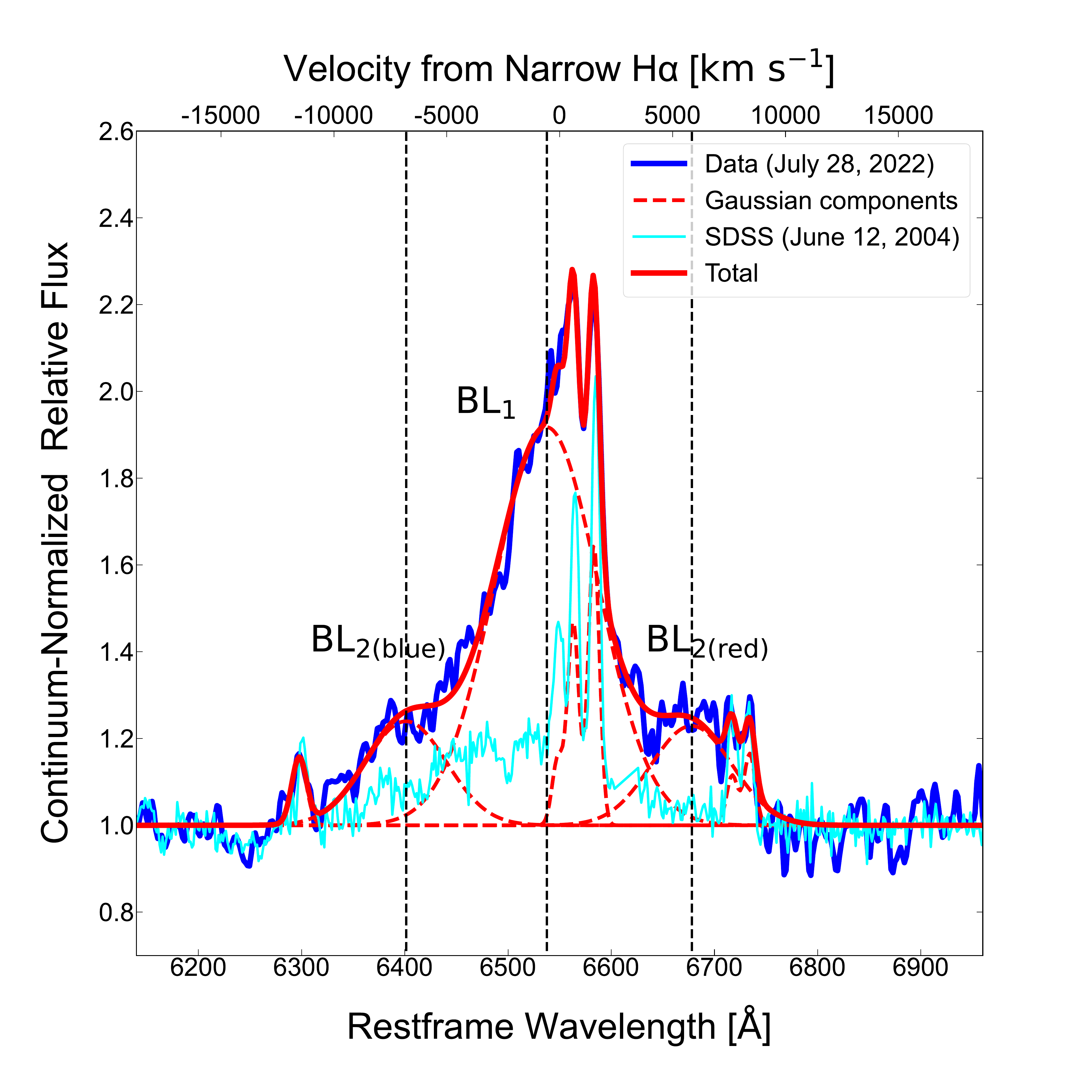}
\caption{The broad $\Ha$ profile of SDSS J1430+2303. The blue line shows the observed spectrum which is normalized by the total continuum on epoch (c). The cyan line shows the continuum-normalized spectrum taken from SDSS on the June 12, 2004. The red dashed and solid lines represent each Gaussian component and the sum of the all models. Narrow $\Ha$ line is set to 0 $\mathrm{km\ s^{-1}}$. The center of $\BLone$ and $\BLtwo$ are plotted by the vertical dashed line. The components $\BLone$ and $\BLtwo$ can be fitted with the velocity width $5\times 10^{3}\ \mathrm{and}\ 4\times 10^{3}\ \mathrm{km\ s^{-1}}$.}
\label{spec}
\end{figure}

The broad $\Ha$ feature can be fitted by the combination of the central component (hereafter referred as $\BLone$) with FWHM $\sim5\times 10^{3}\ \mathrm{km\ s^{-1}}$, which is $\sim 10^3\ \mathrm{km\ s^{-1}}$ blueshifted from the narrow $\Ha$ line, and a symmetrical wing-like pair of the blue and redshifted ($-7\times10^{3},\ +5\times10^{3}\ \mathrm{km\ s^{-1}}$)  components (hereafter referred as $\BLtwo$), as suggested by \citet{jiang2022tick} for the spectra taken in January 2022.
For the component $\BLtwo$, symmetrical profiles with two peaks of the same width and intensity provide reasonable fitting results for the spectrum at each epoch.
The obtained velocity width of the each $\BLtwo$ peak is $\sim4\times 10^{3}\ \mathrm{km\ s^{-1}}$.
We show the best fitting result and a representative continuum-normalized spectrum of \sdss at epoch (c) in Figure \ref{spec} for an example. 

\subsection{Variability of BLR}\label{Variability of BLR}
\begin{figure*}[ht]
\includegraphics[width=170mm]{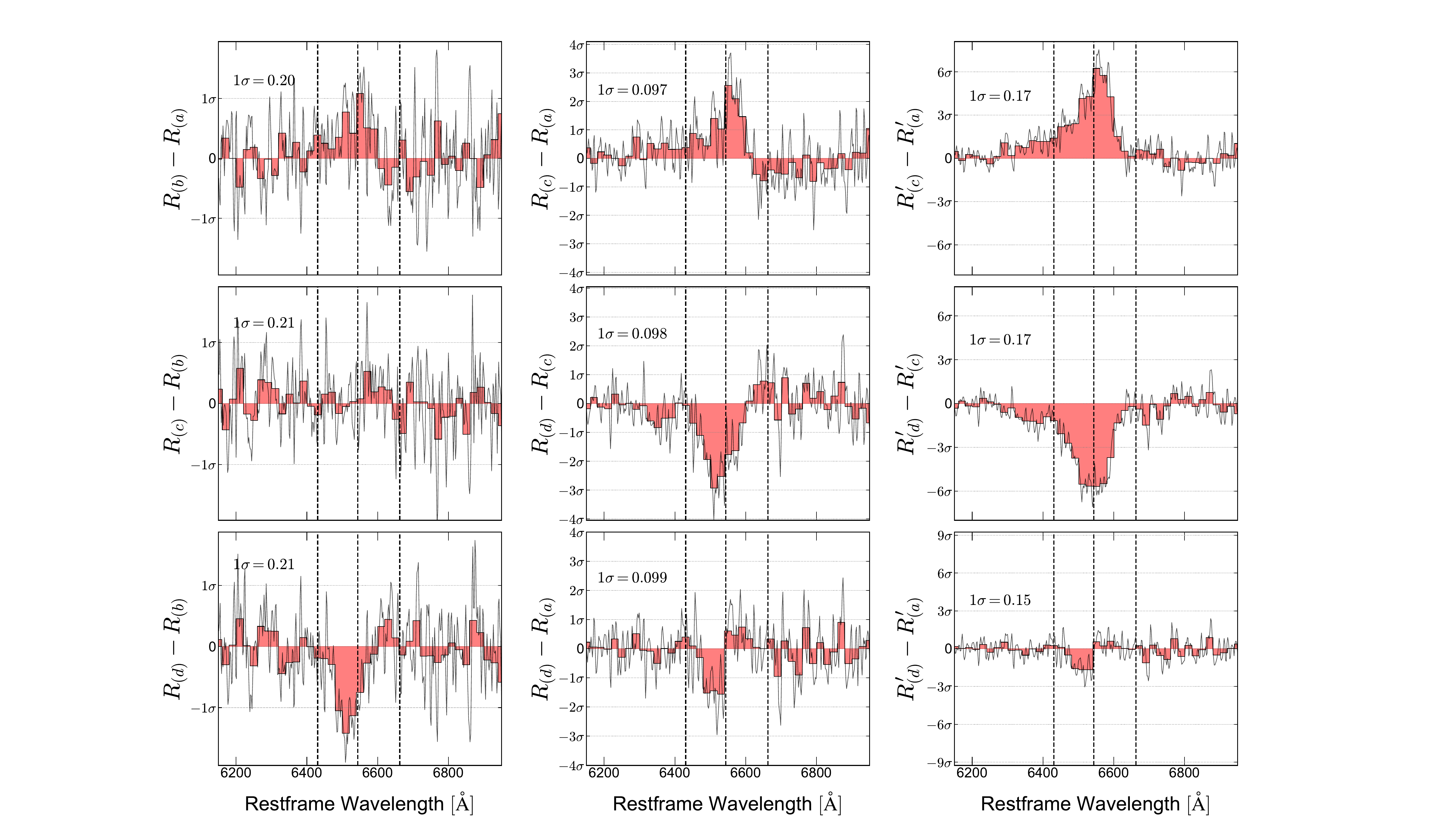}
\caption{(left and middle column): Significance of broad line variations in the continuum-normalized spectra.. Black solid lines show the significance of the difference in the relative flux $R_{(i)}$ and $R_{(j)}$.  $1\sigma$ is a standard deviation of $R_{(i)}$ and $R_{(j)}$ at the wavelength range 6150-6250 and 6850-6950$\AAA$. Each histogram shows the difference in the relative flux $R_{(i)}$ and $R_{(j)}$ with $10\AAA$ bins. The mean values for the center of $\BLone$, $\mathrm{BL}_\mathrm{2(blue)}$ and $\mathrm{BL}_\mathrm{2(red)}$ are plotted by the vertical dashed lines. While each bin does not exceed 3 sigma, the continuous change around the central wavelength of $\BLone$ is significant. The $\BLone$ component exhibits a significant variation around 6550$\AAA$, whereas the $\BLtwo$ component remains invariant. (right column): Significance of broad line variations in the AGN continuum-normalized spectra of $R^{\prime}_{(i)}$.}
\label{rrr40}
\end{figure*}

Figure \ref{rrr40} shows the differences in the relative flux spectra from epoch to epoch. In order to show the significance of the features, we normalized the residuals by the standard deviation values measured at the wavelength ranges used for the continuum fitting. Note that the standard deviations of $R_{(b)}$ in the continuum is 0.20, which is more than twice as large as those of the other epochs (0.08, 0.06 and 0.07 for $R_{(a)}$, $R_{(c)}$, and $R_{(d)}$), since it was obtained using a different higher dispersion grism, VPH683O56. The red histograms represent the residuals in $10\AAA$ bins. 
For comparison, we also show the mean values of the central wavelength of $\BLone$, $\mathrm{BL}_{\mathrm{2(blue)}}$ and $\mathrm{BL}_{\mathrm{2(red)}}$ in all epochs by the vertical dashed lines.

We found significant positive or negative variations in the broad line features ($\BLone$). The relative line flux at around $6550\AAA$ increases from epoch (a) to (b), and (a) to (c) in the top panels of the left and middle column. 
The peak of the variation nearly coincides with the center of $\BLone$.
No significant variation is observed between (b) and (c) in the middle left panel.
The relative flux in epoch (d) is reduced compared to other epochs, leading to negative variations around $6500\AAA$ 

Such variations in the relative flux must occur due to the time lag between AGN continuum variations and the response by the ionized gas. 
Therefore, the central broad component $\BLone$ which shows the variability of the relative flux is considered to be originated from the distant region from the central continuum source such as a BLR.
It is interesting to note that these variations are observed only within the limited wavelength range of the broad $\Ha$ line, implying that the complicated broad emission line feature of \sdss can be physically decomposed into multiple components. 
Given that there is no significant variation in the relative flux around the two peaks of $\mathrm{BL}_{2\mathrm{(blue)}}$ and $\mathrm{BL}_{2\mathrm{(red)}}$ in Figure \ref{rrr40}, it suggests the broad lines respond rapidly to the change of the continuum flux. 
This feature must be originated from the vicinity of the continuum source, significantly closer than a few tens of light days. 
In addition to the variability of the relative flux, we also found that there is a wavelength shift between the peaks of the positive residual from (a) to (c) and the negative residual from (c) to (d). 
This shift can be interpreted as the change of BLR profile responsible for $\BLone$, suggesting that the gas in the region changes its kinematic profile within a yearly observation timescale.
Similar variations are observed in the AGN continuum-normalized spectra after subtracting the host galaxy component, as shown in the right column of Figure \ref{rrr40}.

\sdss is observed in the X-ray and optical wavelength more than 200 days since January 1, 2022.
The light curve in X-ray and optical bands shows a continuous decline from the 156th day, when the large flare occurred, to the 200th day \citep{masterson2023unusual}. 
The increasing of the relative flux $R_{(i)}$ at around the wavelength of $6550\AAA$ from epoch (a) to (b) and (c) is attributed to the continuous decline of the continuum (epoch (a) and (c) correspond to the 146th day and the 208th day after January 1, 2022). 
Quantitatively, the continuum flux in $r$-band, including the host galaxy, decreased by $\sim 10\%$ from epoch (a) to (c) \citep{masterson2023unusual}.
This decrease contributes to the positive residual of the peak relative flux of $\BLone$ by $\sim 0.2$ ($10\%$ of $\sim1.8$ see $\BLone$ component in Fig.\ref{spec}). 
The decrease of $R_{(i)}$ at around the wavelength of $6500\AAA$ from epoch (c) to (d), suggests the recovery of the continuum flux after epoch (c). However it should be noted that no information regarding the light curves in 2023 is available at the time of writing this paper.

We refrain from further speculation on questions such as why the $\BLone$ and $\BLtwo$ structures appeared in the recent spectra in 2022, the origin of the $\BLtwo$, and whether these variations are related to the claimed SMBHB merging event or not. To understand the complexity of the line profile and their physical origins, it is essential to continue monitoring the further variation of this interesting object \sdss.

\begin{ack}
We thank Kyoto University and the National Astronomical Observatory of Japan (NAOJ) for their great help, advice, and cooperation at the Seimei Telescope Observatory.
This work was supported by JST, the establishment of university fellowships towards the creation of science technology innovation, Grant Number JPMJFS2102.
This publication is based upon work supported by KAKENHI (22K03693) through Japan Society for the Promotion of Science.
\end{ack}

\end{document}